\documentclass[12pt]{article}
\usepackage{amsmath,amssymb,amsfonts,amstext}
\usepackage{tabularx}
\usepackage{hyperref}
\usepackage{graphicx}
\usepackage{amsfonts}
\usepackage[usenames]{color}
\usepackage[usenames,dvipsnames,svgnames,table]{xcolor}
\definecolor{Blue}{rgb}{0,0.08,0.45}
\definecolor{Magenta}{cmyk}{0.1,0.8,0,0.1}
\definecolor{Orange}{rgb}{1,0.5,0}

\begin{document}
\title{Discrete graviton spectrum from super-exponential cup potentials and their application to braneworld physics} 
\author{Mariana Carrillo--Gonz\'alez$^{a,b}\!\!$, Gabriel Germ\'an$^{c}$,
Alfredo Herrera--Aguilar$^{d,e}\!\!$,\\
Juan Carlos Hidalgo$^{c}$ and\, Dagoberto Malag\'on--Morej\'on$^{c}$\\
{\normalsize \textit{$^a$Perimeter Institute for Theoretical Physics,}}\\
{\normalsize \textit{31 Caroline St. North, Waterloo ON N2L 2Y5, Canada.}}\\
{\normalsize \textit{$^b$Department of Physics and Astronomy,}} 
{\normalsize \textit{University of Waterloo,}}\\
{\normalsize \textit{200 University Av. West, Waterloo, ON N2L 3G1, Canada}}\\
{\normalsize \textit{$^c$Instituto de Ciencias F\'isicas,} }
{\normalsize \textit{Universidad Nacional Aut\'onoma de M\'exico,}}\\
{\normalsize \textit{Apdo. Postal 48-3, 62251 Cuernavaca, Morelos, M\'{e}xico.}}\\
{\normalsize \textit{$^d$Departamento de F\'{\i}sica, Universidad Aut\'onoma Metropolitana Iztapalapa,} }\\
{\normalsize \textit{San Rafael Atlixco 186, CP 09340, M\'exico D. F., M\'exico.}}\\
{\normalsize \textit{$^e$Instituto de F\'{\i}sica y Matem\'{a}ticas,}}
{\normalsize \textit{Universidad Michoacana de San Nicol\'as de Hidalgo,}}\\
{\normalsize \textit{Edificio C--3, Ciudad Universitaria, CP 58040, Morelia, Michoac\'{a}n, M\'{e}xico.}}}

\date{}
\maketitle

\begin{abstract}

Super-exponential warp factors of the form $e^{-2f} \sim e^{-2c_1e^{c_2 |w|}}$ have been proposed in the context of two-brane models to solve the gauge hierarchy problem of scales by using a compactification scale of the same order of the fundamental Planck scale, completely eliminating the hierarchy between the electroweak and the Planck scales. However, most of the so far studied families of braneworlds with super-exponential warp factors do not localize 4D gravity when one of the branes is sent to infinity. Recently a braneworld model generated by a canonical scalar field $\phi$ minimally coupled to 5D gravity with a bulk cosmological constant was shown to localize 4D gravity and to be stable under linear tensorial and scalar perturbations. Thus, within this model a super-exponential warp factor solves both the hierarchy problem in the two-brane system and the gravity localization one in the single brane configuration. Here we present an explicit new solution for the latter braneworld configuration and study the dynamics of its tensorial perturbations: we find that they obey  a Schr\"odinger--like equation with a well potential which possesses exponentially increasing walls (we call it the {\it cup potential}), yielding a {\it novel discrete spectrum} for the massive Kaluza-Klein (KK) tensorial excitations, despite the non-compact nature of the fifth dimension, and in contrast to the braneworld models proposed so far, where the mass spectrum of the KK modes is continuous, or at most mixed. Namely, the junction conditions on the brane quantize the graviton masses. Finally, the corrections to Newton's law coming from these massive KK modes are analytically computed and estimated. Their novelty arises from the fact that these {\it KK excitations possess quantized masses and are bound to the brane}, however, they are all of the Planck mass order, leading to unreachable energy scales from the phenomenological viewpoint.
 
\end{abstract}

\section{Introduction}

The possible solution of several open problems in modern physics like the cosmological constant problem, the gauge hierarchy problem as well as the localization of  4D gravity based on braneworld models with non-compact extra dimensions has generated considerable interest during last years \cite{rubakov}--\cite{rs}. These models, where our Universe is embedded in a spacetime with extra dimensions, can be classified into thin and thick braneworld configurations. In the former family our Universe is localized in a delta distribution function along the extra dimensions, while in the latter class the branes correspond to smooth bulk configurations in which our world is embedded (for recent reviews see, for instance, \cite{reviews} and references therein). As a straightforward generalization of the original thin braneworlds, there have appeared several models generated with the aid of bulk scalar fields. Different kinds of scalar field configurations coupled to gravity in distinct ways, yield different physical scenarios as it can be seen in e.g. \cite{Cvetic}--\cite{MGAD1}. 

Quite recently a particularly interesting braneworld model generated by a tachyon scalar field minimally coupled to gravity with a super-exponential warp factor $e^{-2f(w)}=e^{-2c_1 e^{c_2 |w|}}$ was successfully used in order to address the gauge hierarchy problem without any fine tuning at all. As a result, scales of the order of the Planck mass are related to the electroweak scale by means of parameters of the same order \cite{koleykar}. However, within these braneworld models the 4D gravity localization condition is never fulfilled unless the tachyonic scalar field is complex \cite{MGAD1}. In general, when the scalar field that generates the braneworld model is a tachyon, attempts to solve the highly non--linear Einstein and field equations give rise to imaginary tachyon field configurations \cite{koleykar}--\cite{palkar}, except for special cases \cite{GADRR}. Notwithstanding, instead of a tachyon braneworld one can use a canonical scalar field coupled to gravity in order to localize 4D gravity with a super-exponential warp factor \cite{MGAD1}.

Here we shall continue to study such a canonical scalar field braneworld model in the presence of a bulk cosmological constant with a super-exponential warp factor with the aim of analyzing the dynamics of the Kaluza--Klein (KK) metric fluctuations. As expected, these perturbations obey a Schr\"odinger--like equation with the peculiarity that the analog quantum mechanical potential now has a {\it cup form} which renders a {\it novel discrete spectrum} of massive metric KK excitations despite the non--compact character of the extra dimension. These massive KK modes yield small violations of unitarity which, in turn, give rise to small corrections to Newton's law. In spite of the mathematical difficulties of this problem, we were able to analytically estimate these corrections. It turns out that these corrections are novel in the sense that they come from a KK graviton spectrum with quantized masses which are bound to the brane, notwithstanding these massive modes are of the Planck mass scale, rendering a hopeless picture from the phenomenological point of view.

The paper is organized as follows: we present the model in Sec. 2 and find a new exact solution to the braneworld field configuration in Sec. 3. We further show that the gauge hierarchy problem can be geometrically reformulated $\mbox{\rm \`a}$ la Randall--Sundrum in Sec. 4. We continue analyzing the dynamics of linear tensorial perturbations when 4D gravity is localized on the brane in Sec. 5, here we also show that the model is stable under these fluctuations and that a discrete spectrum of KK massive modes arises when imposing the junction conditions on the brane. The corresponding corrections to Newton's law are computed and analyzed in Sec. 6 and we summarize our results in Sec. 7.

\section{A braneworld created by a scalar coupled to gravity}

We now consider the action for a canonical scalar field $\phi$ minimally coupled to gravity with a 5D cosmological constant 
$\Lambda_5$ and also introduce a single $3$--brane since we are interested in studying the localization of 4D gravity and the physical implications the extra dimensional world has on our Universe:
\begin{equation}
S\!=\!\int\!d^5x\sqrt{-g}\left[2M_*^3 \left(R\!-\!2\Lambda_5\right)-\frac{1}{2}g^{MN}\nabla_M\phi\nabla_N\phi\!-\!V(\phi)\right]\!-\!\int d^4x\,dw\sqrt{-\tilde g}\,\lambda(\phi)\,\delta(w-w_0) ,
\label{accion2}
\end{equation}
where the 5D gravitational coupling constant is given by $2M_*^3=1/8\pi G_5$, $\lambda$ labels the brane tension, $\tilde g$ is the induced metric on the brane and $w_0$ is the position of the brane along the fifth dimension. The indices take the values $M,N=0,1,2,3,5$. 

The Einstein equations with a bulk cosmological constant and the $3$--brane source are 
\begin{equation}
 G_{MN} + \Lambda_5 g_{MN} 
 = \frac{1}{4M_*^3} \left(T_{MN}^{\it{bulk}}+T_{MN}^{\it{brane}}\right),
\label{einequ2}
\end{equation}
while the Klein-Gordon equation is
\begin{equation}
\Box \phi = - \frac{\partial V_{tot}}{\partial\phi} = \frac{\partial V(\phi)}{\partial\phi} + \frac{\partial\lambda(\phi)}{\partial\phi}\delta(w-w_0).
\label{KGequ}
\end{equation}

The background metric is specified by the ansatz of a warped 5D line element with an induced flat 3--brane that reads
\begin{equation}
ds^2 = e^{-2f(w)} \left[ \eta_{\mu\nu}dx^{\mu}dx^{\nu}+ dw^2  \right].
\label{ansatz}
\end{equation}
The function $f(w)$ is the warp function, while $\mu,\nu$ label the 4D flat spacetime coordinates 
$x^{\mu}$ and we are using the $(-,+,+,+,+)$ signature. 

The energy--momentum tensor in the bulk is given by
\begin{equation}
T_{MN}^{\it{bulk}} =  \nabla_M\phi\nabla_N\phi+
g_{MN}\left (-\frac{1}{2}g^{AB}\nabla_A\phi\nabla_B\phi- V(\phi)\right).
\end{equation}
One can show that the non-diagonal components of the Einstein tensor vanish for the metric ansatz (\ref{ansatz}). Consistency of Einstein equations demands that the non--diagonal components of the overall stress energy tensor should also vanish identically. This fact yields two possibilities: a) the field $\phi$ should depend only on time as in the case of a scalar field in a homogeneous and isotropic background as in cosmology; b) the field $\phi$ depends only on the coordinate
corresponding to the extra dimension, which is the case we shall consider here.

The energy--momentum tensor on the brane can be expressed as follows
\begin{equation}
T_{MN}^{\it{brane}} = - \delta_M^{\mu}\delta_N^{\nu}g_{\mu\nu}\lambda(\phi)\delta(w-w_0).
\end{equation}

The Einstein's equations (\ref{einequ2}) are reduced to a dynamical equation 
\begin{eqnarray}
3\left( f^{''} - f^{'2 }\right)&=& 
\frac{1}{4M_*^3} \left( \frac{1}{2}\phi^{'2} + e^{-2f} V  \right)+\Lambda_5e^{-2f}\, + \frac{1}{4M_*^3} e^{-2f}\lambda\delta(w-w_0),
\label{ein1a}
\end{eqnarray}
and one constraint equation
\begin{eqnarray}
6f^{'2}&=& 
\frac{1}{4M_*^3} \left( \frac{1}{2}\phi^{'2} - e^{-2f} V  \right)-\Lambda_5e^{-2f},
\label{ein1c}
\end{eqnarray}
while the Klein--Gordon equation (\ref{KGequ}) can be written as
\begin{eqnarray}
\phi^{''}-3f^{'}\phi^{'} &=& e^{-2f}\left(\frac{\partial V}{\partial \phi}\,+\frac{\partial\lambda}{\partial\phi}\,\delta(w-w_0)\right),
\label{ecampo1}
\end{eqnarray}
where ${\,^\prime\,}$ denotes derivatives with respect to the extra dimension $w$.

After some algebra from (\ref{ein1a}) and (\ref{ein1c}) we get the following expressions 
\begin{equation}
\phi^{'2}=12M_*^3 \left( f^{''} + f^{'2 }\right) - e^{-2f}\lambda\,\delta(w-w_0),
\label{fip1}
\end{equation}
\begin{equation}
V(\phi)=6M_*^3\left[e^{2f}\left( f^{''} - 3 f^{'2}\right) - \frac{2}{3}\Lambda_5\right] - \frac{1}{2}\lambda\,\delta(w-w_0).
\label{V1}
\end{equation}

These expressions will be very useful when solving the Einstein's equations as well as for studying the structure of the self--interaction potential $V(\phi)$.

\section{A solution with a super--exponential warp factor}

Recently, it has been proposed in \cite{koleykar} a tachyonic braneworld model with a super-exponential warp factor, 
$e^{-2f} = e^{-2c_1e^{c_2 |w|}},$ in order to completely solve the hierarchy problem without further fine tuning, i.e. by setting the fundamental 5D parameters of the model, namely the compactification and Planck scales, to the same order. However, previous attempts to solve the highly non--linear field equations gave rise to imaginary tachyon field configurations which localize 4D gravity, or to real field configurations that do not localize it in the single brane configuration \cite{koleykar}--\cite{GADRR}. Spinor fields were also shown to localize on similar tachyonic braneworlds \cite{zld}. Afterwards, in \cite{MGAD1} it was shown that 4D gravity can be localized if we make use of a canonical scalar field instead of a tachyonic one. Here we shall briefly recall how this result comes about and find a {\it new solution} within this framework. 

Let us begin with a warp function $f(w)$ of the form
\begin{equation}
f \sim c_1 e^{c_2 |w|} \,;
\label{fexp}
\end{equation}
both the finiteness of the effective 4D Planck mass and gravity localization conditions,
\begin{equation} 
0 < \int e^{-3f(w)}dw < \infty, \label{Loca}
\end{equation}
require the constants $c_1$ and $c_2$ to be positive \cite{MGAD1}. 

In order to get the 4D effective Planck scale within our model, we replace the Minkowski metric $\eta_{\mu\nu}$ by a 4D metric $\overline{g}_{\mu\nu}$ in (\ref{ansatz}), a fact which allows us to get an effective 4D Einstein--Hilbert action after integrating over the fifth coordinate $w$ in (\ref{accion2}). We further look at the curvature term from which the scale of gravitational interactions can be derived
\begin{equation}
\label{actioneffective} S_{eff}\supset 2M_{\ast}^3 \int d^4x
\int_{-\infty}^{\infty} dw\sqrt{|\overline{g}|}e^{-3f}\overline{R},
\end{equation}
where now $\overline{R}$ is the 4D Ricci scalar constructed from $\overline{g}_{\mu\nu}$. The effective Planck mass can be easily calculated upon integration with respect to the extra dimension:
\begin{equation}
M_{pl}^2=2M_{\ast}^3\int_{-\infty}^{\infty}dw
e^{-3f}=\frac{4M_{\ast}^3}{c_2}\Gamma\left(0,3c_1\right)=-\frac{4M_{\ast}^3}{c_2}\mbox{\rm Ei}(-3c_1),
\label{Mpl2}
\end{equation}
where the last equality is valid for $c_1>0,\ $ $\Gamma\left(a,x\right)$ is the upper incomplete gamma function and Ei denotes the exponential integral. Thus, this quantity is positive and finite for any positive constants $c_1$ and $c_2$ of our solution, just as required by the gravity localization condition. Therefore, the 4D general relativistic couplings (and those of Newtonian gravity, in particular) are correctly recovered if $M_{\ast}\sim c_2\sim M_{pl}$ and $c_1$ is allowed to vary as 
$10^{-2} < c_1 < 1$.

Thus, $f$ should be a monotonically increasing positive function such that, in the bulk
\begin{equation}
f'^2 \sim c_2^2 f^2 \, , \quad \quad f''\sim c_2^2 f \, > 0 \, ;
\label{fpp}
\end{equation}
we see that $f''$ is also a positive definite function. In general, because of the presence of derivatives of $|r|$ with respect to $r$ in the warp factor, we have $f'' = c_2^2\,f+2\,c_2\, f\,\delta(w-w_0)$, and the presence of the $\delta(w)$ function compensates the introduction of the $3$--brane in the matter sector of the model (\ref{accion2}). 

Therefore, instead of Eqs. (\ref{fip1}) and (\ref{V1}) we now get
\begin{equation}
\phi^{'2}=12M_*^3\left(c_2^2 f+c_2^2 f^2+2\,c_2\, f\,\delta(w-w_0) - \frac{1}{12M_*^3}\,e^{-2f}\,\lambda(\phi) 
\delta(w-w_0)\right),
\label{fip1a}
\end{equation}
\begin{equation}
V(\phi)=6M_*^3\left[e^{2f}\left(c_2^2 f+2\,c_2\, f\,\delta(w-w_0) - 3\,c_2^2 f^2\right) - \frac{2}{3}\Lambda_5
-\frac{1}{12M_*^3}\lambda(\phi) \delta(w-w_0) \right],
\label{V1a}
\end{equation}

Let us now proceed to find a real solution for the scalar field with the super-exponential warp factor (\ref{fexp}).
Denoting by  $\dot F$ a derivative with respect to the absolute value $|w|$ we obtain
\begin{equation}
F''=\ddot F+2 \delta(w-w_0) \dot F\,,
\label{fibar}
\end{equation}
note that $F^{'2}=\dot F^{2}$.
The equation for the field $\phi$ (\ref{ecampo1}) is then
\begin{equation}
\ddot\phi+2\,\dot\phi\,\delta(w-w_0)-3 \dot f\,\dot\phi=e^{-2f}\,\left(\frac{\partial V}{\partial \phi}+
\frac{\partial \lambda}{\partial \phi}\delta(w-w_0)\right).
\label{ecampo2}
\end{equation}
From equations (\ref{fip1a})--(\ref{V1a}) and (\ref{ecampo2}) we see that the brane potential or tension is positive $\lambda|_{w_0}=24M_*^3c_1\,c_2\,e^{2c_1}>0$, while 
$\dot\phi|_{w_0}=\frac{1}{2}\,e^{-2c_1}\,\frac{\partial \lambda}{\partial \phi}|_{w_0}$.  Thus, we obtain for the bulk the following equations
\begin{equation}
\phi^{'2} = 12M_*^3\, c_2^2\,\left(f + f^2\right) ,
\label{fip2}
\end{equation}
\begin{equation}
V(\phi) = 6M_*^3\, c_2^2\,\left[e^{2f}\left(f - 3\,f^{2}\right) - \frac{2}{3\, c_2^2}\Lambda_5\right],
\label{V2}
\end{equation}
with the following solution to Eq. (\ref{fip2}):
\begin{equation}
\phi = \pm \sqrt{12M_*^3}\left(\sqrt{f^2+f}+\mbox{\rm arcsinh}\sqrt{f}\right).
\label{fisol}
\end{equation}

It is straightforward to check that the super-exponential warp factor (\ref{fexp}) and the scalar field (\ref{fisol}) identically satisfy the field equations (\ref{ein1a})-(\ref{ecampo1}) in the bulk.

\begin{figure}[htb]
\begin{center}
\includegraphics[width=8cm]{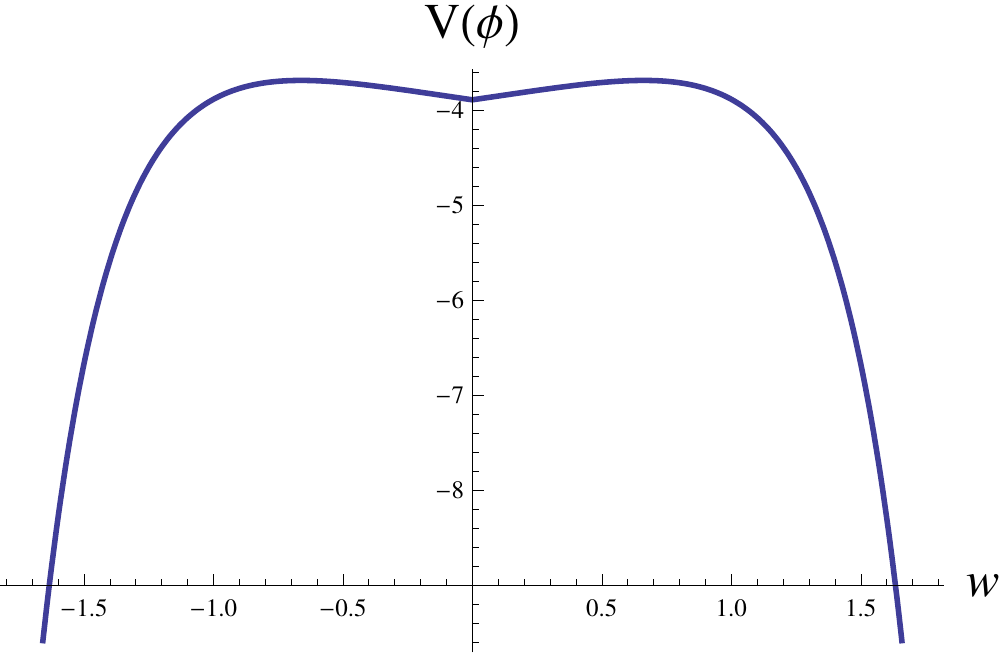}
\end{center}
\caption{The self--interaction potential $V$ as a fuction of the coordinate w. The values used for the parameters are $c_1=0.3$, $c_2=1$ and $M_*=1$.}
\label{c}
\end{figure}

The profile of the self--interaction potential $V(w)$ can be obtained by means of \eqref{fexp} 
and (\ref{V2}). This is presented in Fig. \ref{c}.
This potential is unbounded from bellow and seems to be pathological. However, this kind of potentials is very common and they are stable in AdS supergravity with one scalar field as it was shown on the basis of the positive energy theorem in \cite{townsend} (see references therein as well). The stability of the scalar--tensor braneworld system under consideration was analyzed in detailed in \cite{bhknq} for AdS$_5$ backgrounds with a self--interaction potential very similar to (\ref{V2}).

\section{Solution to the gauge hierarchy problem}

Let us approach the gauge hierarchy problem within the framework of the braneworld model under consideration. We now need two branes: the so--called TeV brane, where the SM particles are trapped, located some distance away in the $w$--direction from the Planck brane, where 4D gravity is localized. Thus, an extra brane is needed so that the second integral in the action (\ref{accion2}) should be replaced by one of the form
\begin{equation}
S_B = -\sum_{i=1}^2 \int d^4 x d w \sqrt{-\tilde g_i} \lambda_i(\phi) \delta(w-w_i)\,,
\label{actionB}
\end{equation}
where $\tilde g_i$ and $\lambda_i$ respectively are the induced 4D metric and the brane tension corresponding to the i-th brane located at $w_i$.

When the TeV probe brane is located at the position
$w_0$ along the extra dimension, the SM particles feel the effective 4D metric:
\begin{equation}
g_{\mu\nu}^{SM}=e^{-2f(w_0)}\overline{g}_{\mu\nu}=e^{-2c_1e^{c_2|w_0|}}\overline{g}_{\mu\nu}.
\end{equation} 
Thus, the physical mass scales are set by the symmetry--breaking scale \cite{rs}
\begin{equation} m=e^{-c_1e^{c_2|w_0|}}\ m_0.
\end{equation}
Therefore, in order to produce TeV physical mass scales from fundamental Planck mass
parameters, the following relation must hold 
\begin{equation}
\frac{TeV}{M_{pl}}
\approx e^{-c_1e^{c_2|w_0|}},
\end{equation} 
a condition which yields, for instance, $c_1=1$ and $c_2w_0\approx 3.6$ (or $c_1=2$ and $c_2w_0\approx2.9$, or, respectively, $c_1=3$ and $c_2w_0\approx2.5$), leading to a setup in which there is no hierarchy at all between the parameters of the model: $c_1\sim c_2\sim w_0\sim 1$. In view of this fact, it is worth mentioning that within the model with two branes, it is possible to completely ignore the numerical parameter $c_1$ since it can be set to unity $c_1=1$.

It is worth mentioning that the tension on the TeV brane of the super--exponential two--brane model is different from the tension on the Planck brane, both in magnitude as well as in sign, due to the modified character of the warp factor compared to the Randall--Sundrum setup. However, as well as in this latter model, both tension branes must be fine--tuned in order to set effective 4D cosmological constant to zero. Both of these results were explicitly shown in \cite{koleykar}.

We should address as well the stability problem of the brane separation $w_0$ between the two 3--branes. As in the Randall--Sundrum model, the introduction of the TeV brane some distance away from the Planck (gravitational) brane, in order to obtain the desired warping from the Planck scale to the TeV energy scale and hence the needed hierarchy, gives rise to a fine--tuning on the brane position $w_0$. Therefore, we need to stabilize this brane separation. This aim can be achieved by means of the so--called Goldberger--Wise mechanism by associating to the brane separation a scalar field, with specific interaction terms on each brane, that models the size of the fifth dimension. This study was performed first neglecting the back--reaction of the TeV brane in \cite{GW} and it was further generalized to the case with back--reaction in \cite{dewolfe}. In the latter case it was shown that the brane separation $w_0$ remains stable when modeled by a scalar field with quartic self--interaction potentials.

\section{Tensor perturbations and discrete graviton spectrum}
\label{tensor}

In this Section we shall briefly review the localization of 4D gravity in our braneworld field configuration following the work presented in \cite{Giovannini}, where the stability of metric fluctuations of higher--dimensional backgrounds with non--compact extra dimensions was accomplished and the master equations for the coupled system of metric and scalar perturbations were also derived in a gauge--invariant form. 

Let us start with the following perturbed ansatz in order to analyze the dynamics of the metric fluctuations 
\begin{equation}
ds^2 = e^{-2f} \left[\left(\eta_{\mu\nu} + h_{\mu\nu}\right)dx^\mu dx^\nu + dw^2\right],
\label{metricpert}
\end{equation}
where $h_{\mu\nu}(x^\mu,w)$ are gauge--invariant metric perturbations when restricting to the transverse and traceless sector $\partial^\mu h_{\mu\nu}=h^\mu_\mu=0$. By making further use of the separation of variables $h_{\mu\nu} = C_{\mu\nu} e^{3f/2} e^{ipx} \rho(w)$, where $C_{\mu\nu}$ are arbitrary constants, the dynamical equation along the extra dimension adopts a Schr\"odinger--like form
\begin{equation}
- \rho'' + V_{QM}(w) \rho = m^2 \rho, 
\label{Schreqnrho}
\end{equation}
where $m$ is the mass that a 4D observer measures, whereas the analog quantum mechanical potential 
reads
\begin{equation}
V_{QM}(w) = \frac{s''}{s} \equiv {\cal J}^2 - {\cal J}', \qquad \mbox{\rm with} \qquad s = e^{-3f/2}.
\label{potVQMrho}
\end{equation}
The quantity ${\cal J} = -\frac{s'}{s}$ is called {\it superpotential} within the context of supersymmetric quantum mechanics since it allows us to write the Schr\"odinger equation (\ref{Schreqnrho}) in the following form
\begin{equation} 
{\cal Q}^{\dagger} {\cal Q}\ \rho = m^2 \rho,
\label{susyqmrho}
\end{equation}
where the operators ${\cal Q}^{\dagger}$ and ${\cal Q}$ are defined as
\begin{equation}
{\cal Q}^{\dagger} = \biggl( - \frac{d}{dw} + {\cal J} \biggr), \qquad\qquad
{\cal Q} = \biggl( \frac{d}{dw} + {\cal J} \biggr).
\label{operators}
\end{equation}
The hermiticity and positive definite character of the left hand side of  (\ref{susyqmrho}) guarantee a positive spectrum of metric fluctuations, evidencing the lack of tachyonic modes with $m^2 <0$ and ensuring the stability of the system under the tensorial sector of perturbations. 

It is remarkable that the massless zero mode of the Schr\"odinger--like equation (\ref{Schreqnrho}) yields $\rho=s=e^{-3f/2}$ and the 
normalization condition (\ref{Loca}) is fulfilled. This integral converges for positive $c_1$ and $c_2$, guaranteeing the 
existence of a normalizable massless zero mode which is interpreted as the 4D graviton localized on the brane.

It is worth noticing that in \cite{Giovannini} it was also shown that the scalar--tensor coupled system (\ref{accion2}) that defines our braneworld is stable under linear scalar fluctuations and free of  tachyonic modes with $m^2 <0$ thanks to the fact that the corresponding Schr\"odinger equations can also be written in the form (\ref{susyqmrho}) for the scalar sector of 
perturbations (see as well \cite{KKS}, \cite{townsend} and \cite{mg} for an exhaustive study of scalar fluctuations in AdS$_5$ spacetimes). 

Moreover, in \cite{MGAD1} it was shown that the involved quantum mechanical potentials are positive barriers distributed along the fifth dimension and no massless nor massive scalar modes are localized on the brane generated by a super--exponential warp factor. 

\begin{figure}[htb]
\begin{center}
\includegraphics[width=8cm]{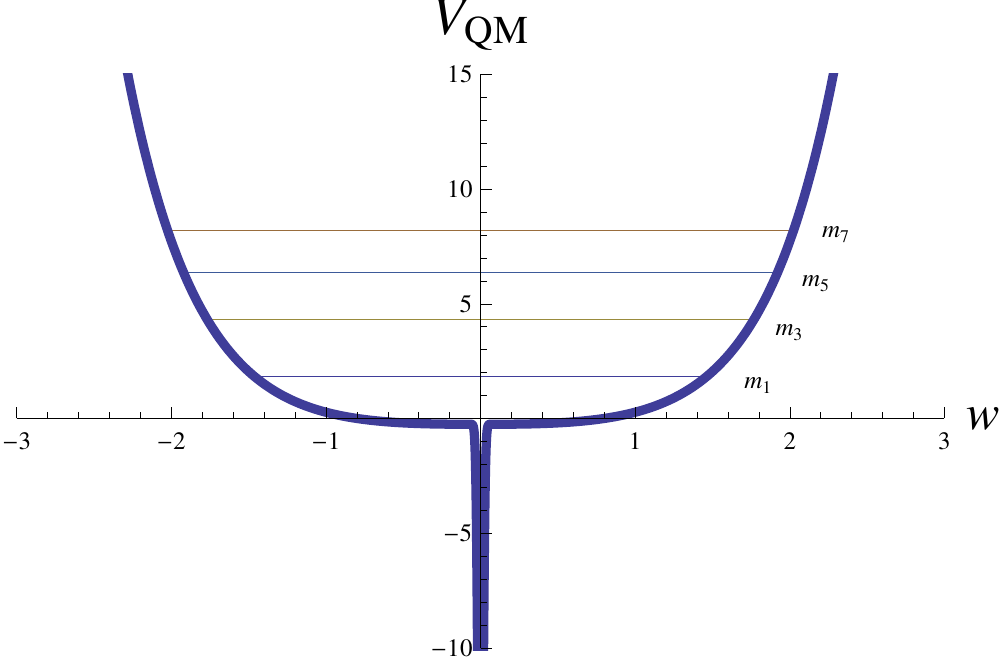}
\end{center}
\caption{The quantum mechanical potential $V_{QM}(w)$ of Eq.~\eqref{VQMw} presents a cup profile.  In this figure we set $c_{2}=1$ and $c_1 = 0.3$. The thin lines corresponds to the first few odd massive states.}
\label{cup}
\end{figure}
Let us return to the analog quantum mechanical potential (\ref{potVQMrho}), which in terms of the coordinate $w$ reads
\begin{align}
V_{QM}(w) & =- \frac{3}{2}c_2^2f+\frac{9}{4}c_2^2f^2-3c_2f\delta(w) \notag \\
&=\frac{3}{2}c_2^2c_1\left(\frac{3}{2}c_1e^{2c_2|w|}-e^{c_2|w|}\right)-3c_1c_2\delta(w)
\label{VQMw}
\end{align}

\noindent and resembles a {\it cup potential} as it can be appreciated from Fig. \ref{cup}. From the form of the potential we see that the spectrum of massive KK excitations will be discrete since its walls are infinite and that the massless zero mode will be a bound state localized on the brane, a fact guaranteed by the existence of the negative delta function at the 3--brane position.

The general solution to Eq. \eqref{Schreqnrho} is given by
\begin{equation}
\rho(w)=e^{-\frac{3}{2}f} f^\alpha \left[k_1 U\left(\alpha,2\alpha+1,3f \right)+ k_2 L_{-\alpha}^{2\alpha}\left(3f \right)\right],
\label{soln}
\end{equation}
where $\alpha\equiv i\frac{m}{c_2} $ and $k_1, k_2$ are integration constants. $U\left(a,b,x\right)$ is the confluent hypergeometric function and 
$L_{n}^{a}\left(x\right)$ denote generalized Laguerre polynomials. However, it can be shown that the latter are singular for any value of the mass parameter $m$ and therefore we shall set $k_2=0$. Hence, we shall consider the term containing the confluent hypergeometric function as the solution to the Schr\"odinger equation (\ref{Schreqnrho}).

\begin{figure}[htb]
\begin{center}
\includegraphics[width=8cm]{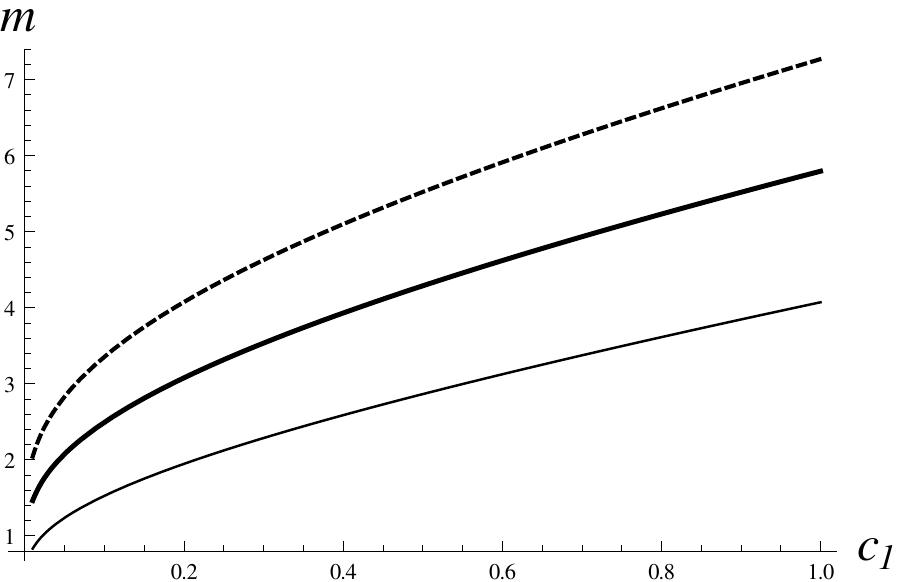}
\end{center}
\caption{The first three eigenmasses of the KK spectrum of excitations in Planck mass units. In this figure we set $c_{2}=1$ and
$c_1$ varies from $.01$ to $1.$ The thin line corresponds to the first massive state, the thick one to the second state 
and the dashed line to the third one.}
\label{spectrum}
\end{figure}

The junctions conditions are obtained by integrating \eqref{Schreqnrho} with the potential (\ref{VQMw}) over a small domain $\epsilon$ around the brane, a procedure which yields
\begin{equation}
-\int\limits_{-\epsilon}^{\epsilon}\rho''(w) dw+\int\limits_{-\epsilon}^{\epsilon} V_{QM}(w)\rho(w) dw=m^2\int\limits_{-\epsilon}^{\epsilon}\rho(w) dw,
\end{equation}
and taking the limit $ \epsilon\rightarrow 0 $ from the left and right sides (denoted by the $-$ and $+$ subindices, respectively): 
\begin{equation}
\rho'_+(0)-\rho'_-(0)=\lim\limits_{\epsilon\rightarrow 0}\int\limits_{-\epsilon}^{\epsilon} (-3c_1c_2e^{c_2|w|})\rho(w)\delta(w) dw=-3c_1c_2\rho(0).
\label{jc}
\end{equation}
We further compute the derivatives of \eqref{soln} from the left and right sides (with $k_2=0$) and substitute them into \eqref{jc} to get 
\begin{align}
3c_1U\left(\alpha+1,2\alpha+2,3c_1  \right)=U\left(\alpha,2\alpha+1,3c_1  \right). 
\label{cu}
\end{align}
By making use of eq. (13.4.18) of \cite{Abramowitz} 
\begin{equation}
(b-a)U(a,b,z)+U(a-1,b,z)-zU(a,b+1,z)=0,
\end{equation}
we get the following result  
\begin{equation}
\frac{m}{c_2}\,U\left(\frac{i m}{c_2}+1,\frac{2 i m}{c_2}+1,3c_1  \right)=0.
\label{quantmasses}
\end{equation} 
This product quantizes the 4D masses of the graviton spectrum of 
KK excitations of our braneworld model in Planck mass units since 
$c_2$ is of that order if we want to obtain the correct gravitational couplings 
of the massless zero mode that parameterizes the 4D graviton. In fact, 
the first factor yields the massless zero mode, while the zeroes of the 
confluent hypergeometric function define the discrete massive bound 
states of the system. In Fig. \ref{spectrum} 
we show the behaviour of the first massive modes with respect to $c_1.\!\!\!$
\footnote{These calculations were done numerically.}

From this figure we see that for all the range of $c_1$ the
mass gap is of Planck mass order since the lightest 
KK massive mode bounded to the 3--brane will be excited 
at this energy scale. The reader should note that this behaviour is not surprising due to the nature of our spectrum: If the spectrum is discrete, the KK masses are generically determined by the coupling constants of the fundamental theory. Thus, when we demand no hierarchies among the coupling constants, the values of the KK masses are always close to the Planck scale.

In the next section we compute the 
corresponding corrections to the Newton's law in the weak field 
limit. This eigenvalue problem is interesting enough in its own right since it renders a 
discrete spectrum of masses (as in the KK setup) even when 
the extra dimension is non--compact.

\section{Weak field limit and corrections to Newton's law}

Let us consider the gravitational potential between two test bodies, with masses $M_1$ and $M_2$, bound to the brane and separated by a distance $r$ in the weak field limit. It turns out that all the discrete massive gravitons give rise to small corrections to Newton's law in ordinary 4D flat spacetime. These corrections can be expressed as an infinite sum due to the discrete character of the spectrum \cite{bs}:
\begin{eqnarray}
V(r)\sim \frac{M_1M_2}{r} \biggl( G_4 + M_{\ast}^{-3} \sum_{j=1}^\infty e^{-m_jr} \bigr\vert\rho_j(0)\bigl\vert^2 \biggr)=\frac{M_1M_2}{r} \big( G_4 + \Delta G_4 \big). 
\label{V}
\end{eqnarray}
where the position of the brane, which represents our physical universe, was set to $w_0=0$ for simplicity, $G_4$ denotes the four dimensional
gravitational coupling, and $\rho_j$ are independent eigenfunctions corresponding to the spectrum of normalizable massive excitations $m_j$ under the sum.

From (\ref{soln}) it follows that the eigenfunctions on the brane read
\begin{eqnarray}
\rho_{j}(0)= k_{j}(m_j)c_1^{im_j/c_2}e^{-3c_1/2}U\Bigl(\frac{im_j}{c_2},1+\frac{2im_j}{c_2},3c_1 \Bigr), 
\label{rhoi}
\end{eqnarray}
where $k_j$ stand for the normalization constants corresponding to each bound state with mass $m_j$. 

Thus, we first need to compute the normalization constants $k_j$ from the following expression 
\begin{equation}
\int_{-\infty}^\infty \bigr\vert \rho_j(w) \bigl\vert^{2} dw = \frac{2}{c_2}\int_{3c_1}^\infty \frac{e^{-z}}{z} \bigr\vert U(\alpha_j,2\alpha_j+1,z) \bigl\vert^{2} dz=1,
\label{normalizn}
\end{equation}
where we have introduced the new variable $z=3f=3c_1e^{c_2\vert w\vert}$. After 
performing this integral numerically, we can easily compute the normalization 
constants $k_j$ for each $m_j$. 
Therefore, the final expression for the corrections to Newton's law $\Delta G_4$ reads
\begin{eqnarray}
\Delta G_4\!=\!M_{\ast}^{-3} \sum_{j} e^{-m_jr} \bigr\vert\rho_j(0)\bigl\vert^2.
\label{correctns}
\end{eqnarray}
\begin{figure}[htb]
\begin{center}
\includegraphics[width=8cm]{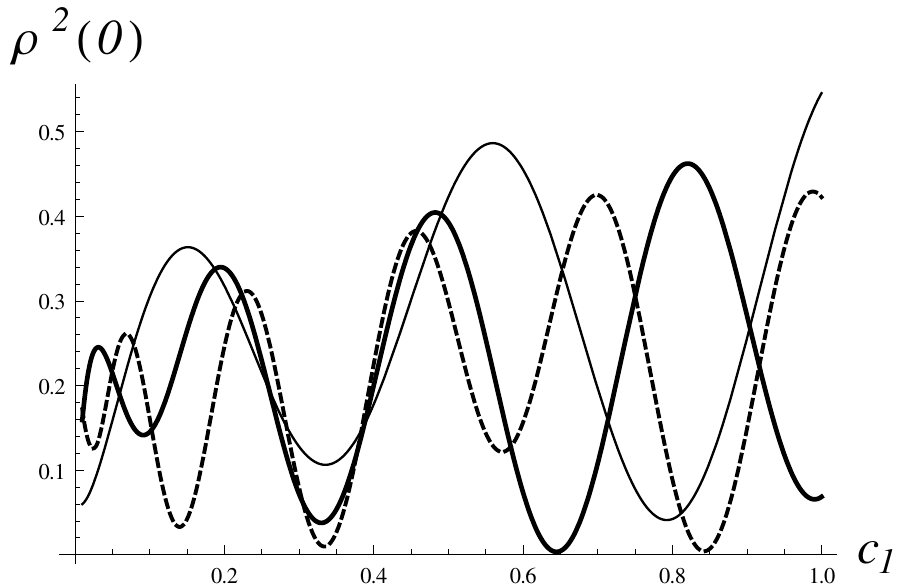}
\end{center}
\caption{The profile of $\rho^{2}(0)$ associated to the first three eigenvalues of the massive KK spectrum in Planck mass units. 
Here $c_{2}=1$ and $c_1$ varies from $.01$ to $1.$ The thin line corresponds to the first massive state, the thick one to the second state 
and the dashed line to the third one.}
\label{rho0vsc1}
\end{figure}

Here we should estimate the contribution of the first KK excitations since the prefactor $e^{-m_jr}$ make them decay very fast. In fact, since the masses $m_j$ are of the Planck scale order, even the second bound state contributes much less than the first state, the third bound state contributes much less than 
the second one, and so on. However, we should also take into account the prefactor $\bigr\vert\rho_j(0)\bigl\vert^2$ in 
order to make a complete analysis. In Fig. \ref{rho0vsc1} we present the profile of $\bigr\vert\rho_j(0)\bigl\vert^2$ vs the $c_1$ 
constant for the first three KK massive excitations. From these graphs we observe that, for the considered range 
of the constant $c_1$, the contribution of the prefactor $\bigr\vert\rho_j(0)\bigl\vert^2$ to the summation 
in \eqref{correctns} is of the same order, so that the summation as a whole is dominated by the exponential 
prefactor. Thus, even if we consider the first three eigenmasses for simplicity, the summation will be 
strongly dominated by the first excited eigenstate, which leads to highly suppressed corrections, 
unreachable to present day experiments but possibly observable in cosmological probes as we discuss in the following section.

\section{Summary of results and discussion}

In this paper we have upheld the idea that super-exponential warp factors provide the possibility of addressing the hierarchy problem in braneworld models with fundamental parameters of the same order, resolving the gauge hierarchy without any fine--tuning. A similar result was obtained for the first time in \cite{koleykar} on the basis of a tachyonic scalar field braneworld model. However, this kind of warp factors usually does not lead to localization of 4D gravity when making use of tachyonic scalar fields in order to model the extra dimension \cite{MGAD1}.

Here we have studied a thin super-exponential braneworld model with a canonical scalar field minimally coupled to gravity in the presence of a 5D cosmological constant as the main ingredients. It turns out that despite the technical difficulties of working with super--exponential warp factors, it is possible to analytically solve the involved field equations. Thus, we have found a new scalar field configuration for which both the 4D gravity localization and the gauge hierarchy problems can be tackled with the same kind of warp factors, contrary to previous results where the scalar field usually turns out to be complex when attempting to localize 4D gravity in the $3$--brane (see, for instance, \cite{koleykar}--\cite{MGAD1}). 

The obtained exact field configuration is stable and allows us to study the dynamics of tensorial fluctuations of the metric. These perturbations obey a Schr\"odinger--like equation with an analog quantum mechanical potential that resembles a {\it cup potential} with exponentially increasing walls and renders a {\it novel discrete spectrum} of massive KK modes in contrast to previous results reported in the literature (to the best of our knowledge). By analytically approaching the corresponding eigenvalue problem, we exactly solved the Schr\"odinger--like equation in terms of confluent hypergeometric functions with complex parameters. In fact, the junction condition on the brane (defined by the zeroes of a confluent hypergeometric function) quantizes the graviton mass spectrum. The scale separation determined by the mass gap in this spectrum is of the order of the Planck mass, yielding energy scales far bigger than the TeV scale which will not be reached in near future experiments.

The corresponding corrections to Newton's law coming from the extra dimensional massive spectrum of KK metric excitations were estimated as well. These corrections are extremely small since they exponentially decay due to the presence of the above mentioned mass gap, which is of the order of the Planck energy scale. Consequently, a 4D observer living in the TeV 3--brane will not be able to detect those KK excitation modes nowadays. 

While weak field effects are unobservable, it is evident that the effects of the studied braneworld will be most prominent in high energy environments. This motivates the study of cosmological spacetimes where the corrections to the gravitational interaction affect significantly the inflationary universe. 
This will yield observable effects from the early stages of evolution of the universe like the evaporation of primordial black holes \cite{liddle}, and most importantily the modifications to the spectrum of primordial gravitational waves \cite{BouhmadiLopez:2004ax,Okada:2014nia}.

We finally would like to stress that the eigenvalue problem that arose when studying the dynamics of the tensorial sector of metric perturbations is quite interesting by itself thanks to the discrete nature of the massive spectrum that it yields even when the extra dimension is non--compact. Recall that in the original KK setup the higher dimensions are compact.

\section*{Acknowledgements}

We gratefully acknowledge support from \textquotedblleft Programa de Apoyo a Proyectos de Investigaci\'on e Innovaci\'on Tecnol\'ogica\textquotedblright\, (PAPIIT) UNAM, IN103413-3, {\it Teor\'ias de Kaluza--Klein, inflaci\'on y perturbaciones gravitacionales.} AHA is heartily grateful to the staff of ICF, UNAM and the Physics Department of UAM-I for hospitality. DMM acknowledges a postdoctoral grant from DGAPA--UNAM. MCG acknowledges PI and CONACYT scholarships provided to undertake graduate studies. GG, AHA, JCH and DMM thank SNI--CONACYT for support.

\end{document}